# Recent progress in perpendicularly magnetized Mn-based binary alloy films

Zhu Li-Jun, Nie Shuai-Hua, and Zhao Jian-Hua[†]

*State Key Laboratory of Superlattices and Microstructures, Institute of Semiconductors, Chinese Academy of Sciences, P. O. Box 912, Beijing 100083, China*

In this article, we review the recent progress in growth, structural characterization, magnetic properties and related spintronic devices of tetragonal $Mn_xGa$ and $Mn_xAl$ thin films with perpendicular magnetic anisotropy. In the first part of this review, we present a brief introduction to the demands for perpendicularly magnetized materials in spintronics, magnetic recording and permanent magnets applications, and the most promising candidates of tetragonal $Mn_xGa$ and $Mn_xAl$ with strong perpendicular anisotropy. Then, in the second and third parts, we focus on the recent progress of perpendicularly magnetized $Mn_xGa$ and $Mn_xAl$, respectively, including their lattice structures, bulk synthesis, epitaxial growth, structural chracterizations, magnetic and other spin-dependent properties, and spintronic devices like magnetic tunneling junctions, spin valves and spin injector into semiconductors. Finally, we give a summary and a perspective of these perpendicularly magnetized Mn-based binary alloy films for future applications.

## 1. Introduction

Ferromagnetic films with high perpendicular magnetic anisotropy ($K_u$) have attracted much attention for their potential applications in nanoscale magnetic devices with high thermal stability.[1-4] Spin valves and magnetic tunneling junctions (MTJs) with high $K_u$ have been demonstrated to allow for realization of high-sensitivity magnetoresistive sensors, spin-transfer-torque (STT) switching Gbit-class magnetoresistive random access memory (MRAM) and high-power oscillators.[2-5] Magnetic materials which have high $K_u$ along with high spin polarization ($P$), high Curie temperature ($T_c$), flexible magnetization ($M_s$) and low magnetic damping constant (α) are the most ideal combination for spin-transfer-torque applications to facilitate reliable switching under the current density below $10^5$ $A/cm^2$.[2-4] Moreover, ferromagnetic films with both high $K_u$ and good compatibility with semiconductor allow for not only the development of ferromagnetic metal/semiconductor hybrid devices including spin field effect transistor,[6,7] spin Hall transistor,[8] light emitting diodes[9,10] and lateral spin valves[12] with perpendicular anisotropy, but also high-density integration of metallic spintronics functional devices like nonvolatile magneto-resistive random access memory on semiconductor photonic and electronic circuits.[12,13-17] These approaches could also perform logic, communications and storage within the same materials technology taking advantage of nonvolatility, increased integration density, increased speed and decreased energy consumption compared with conventional charge-based devices.[18] Furthermore, demand for increasing high-density data storage leads to the replacement of in-plane magnetic recording by perpendicular magnetic recording. Perpendicular magnetic recording with areal density beyond 10 $Tb/inch^2$ requires films with high $K_u$ up to $10^7$ erg/cc and moderate magnetization $M_s$.[19] Too high $M_s$ may result in strong coupling between adjacent bits, while too low $M_s$ brings about low signal-noise ratio. Besides, rare-earth magnets provide the backbone of many products including computers, mobile phones, electric cars and wind-power generators. However, because of both the increasing limited availability and high costs of mining and processing of rare-earth elements, new kind of rare-earth-free magnets are urgently needed.[16, 20] Permanent magnets applications require materials that have large magnetic energy product, high intrinsic and extrinsic coercivities and linear demagnetization curve in the second quadruple.

As the most outstanding representatives, perpendicularly magnetized Mn-binary alloys $Mn_xGa$ (MnGa) and $Mn_xAl$ (MnAl) have attracted much attention and have been studied intensively in the past couple of decades. $L1_0$-MnGa ($D0_{22}$-$Mn_3Ga$) alloys are theoretically expected to have $K_u$ of 26 (20) Merg/cc,[21,22] $M_s$ of 845 (305) emu/cc,[16,21-23] $(BH)_{max}=(2\pi M_s)^2$ of 28 (3.7) MGOe,[17,24] $P$ of 71% (88%) at the Fermi level [17,21,24,25] and α of 0.0003 (0.001),[22] respectively. Moreover, $L1_0$-MnAl are theoretically predicted to have a $K_u$ of 15 Merg/cc, $M_s$ of 800 emu/cc (or 2.37 $\mu_B$/Mn) and $(BH)_{max}=(2\pi M_s)^2$ of 12.64 MGOe.[26,27] These fascinating magnetic properties make MnGa and MnAl promising in spintronics, ultrahigh-density magnetic recording and economical permanent magnets applications. Moreover, many experimental attempts have been made to the preparation and characterization of MnGa and MnAl bulks and films, and their related spintronic devices like magnetic tunneling junctions, spin valve and spin-injection devices. In this article, we will present an overview of recent progress on the growth，structural characterization, magnetic properties and some other properties of perpendicularly magnetized $Mn_xGa$ and $Mn_xAl$ alloys and their related spintronic devices.

---
* Project supported by NSFC 11127406
† Corresponding author: jhzhao@red.semi.ac.cn

## 2. Mn$_x$Ga with perpendicular magnetic anisotropy
### 2.1 Lattice structure and synthesis of Mn$_x$Ga bulks

Alloys of Manganese (Mn) and Gallium (Ga) have quite complicated phase diagram (Fig.1a) with several magnetically ordered phases.[28-30] This review is mainly confined to two most interesting tetragonal phases with strong magnetism and high Curie temperature: the first is $L1_0$-ordered ($\gamma_3$) thermal-dynamically ferromagnetic phase for $0.76 \leq x \leq 1.8$;[17,31,32] the second is $D0_{22}$-ordered ($\varepsilon$) ferrimagnetic phase, forming when $2 \leq x <3$.[25,33-36] The lattice units of both $L1_0$-MnGa and $D0_{22}$-Mn$_3$Ga are outlined in Fig. 1(b). The lattice unit of $L1_0$-MnGa consists of Mn and Ga monatomic layers stacked alternately along the $c$ axis. The lattice constants of $L1_0$-MnGa bulk (space group $P4/mmm$) are $a$=3.88-3.90 Å and $c$=3.64-3.69 Å.[31,32] Bither et al. synthesized $L1_0$-Mn$_x$Ga ($1.2<x \leq 1.5$) polycrystalline bulks.[31] Lu et al. systemically investigated the crystalline structures of Mn$_x$Ga bulks in a broad composition range of $1.15<x<1.89$.[32] For perfect stoichiometric $L1_0$-MnGa ($x$=1), each Mn atom is expected to contribute a magnetic moment of 2.51 $\mu_B$ (corresponding to magnetization of 845 emu/cc); the magnetism of $L1_0$-Mn$_x$Ga may also strongly be influenced by compositions and strains.[16,17,23,25,31,36] As theoretically predicted, excess Mn atoms will align antiparallel to the rest and result in compensation, and strains will also induce a decrease of Mn moment.[23,31] For $D0_{22}$-Mn$_x$Ga bulk (space group $I4/mmm$), the lattice constants are $a$=3.90-3.94 Å and $c$=7.10-7.17 Å.[33-35] Krén and Kádár obtained $D0_{22}$-Mn$_{2.5}$Ga bulks by annealing hexagonal $D0_{19}$-Mn$_x$Ga at 350-400 $^o$C for 1-2 weeks, and revealed the ferrimagnetism with Mn atom magnetic moment of 2.8 ± 0.3 $\mu_B$ and 1.6 ± 0.2 $\mu_B$ at site I and site II, respectively, by neutron diffraction measurement.[33] Niida et al. and Winterlik et al. synthesized polycrystalline $D0_{22}$-Mn$_x$Ga bulks in the composition range of $1.94 \leq x \leq 2.85$[34] and $2 \leq x \leq 3$,[25] respectively, and revealed the composition-sensitive magnetization in the magnetically frustrated ferrimagnets. With Mn content increasing, lattice constant $a$ slightly increases or keeps constant, while $c$ always decreases monotonically.[25, 31-36] Huh et al. synthesized Mn$_x$Ga nano-ribbons with $L1_0$ ($x$=1.2, 1.4, 1.6) and $D0_{22}$-ordering ($x$=1.9) using arc-melting and melt-spinning followed by a heat treatment and investigated their magnetic and transport behaviors.[37] The Curie temperature of Mn$_x$Ga nano-ribbons are found to increase with increasing $x$. All the ribbons are metallic transport behavior, in consistent with MnGa films.

The magnetic properties of Mn$_x$Ga alloys, especially giant perpendicular anisotropy, high spin polarization, high Curie temperature, flexible magnetization and low magnetic damping constant are promising for applications of magnetic recording, permanent magnets and spintronics like magnetic sensors, memories and logics.[16,17] Reliable growth of smooth MnGa thin films and effective tailoring of the magnetism are of particular importance for the possible practical applications in not only metal spintronics but also semiconductor spintronics. In the following, we will overview the recent progress in the growth and characterizations of $L1_0$-MnGa films.

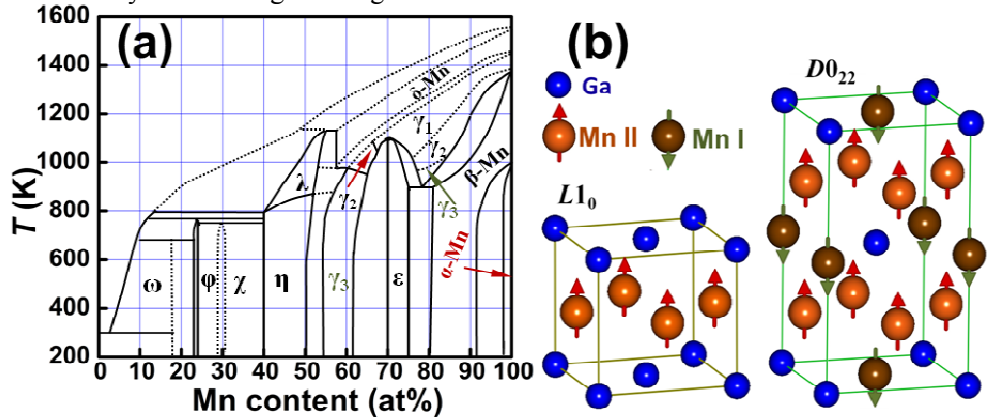

Fig. 1. (a) Phase diagram of Mn-Ga alloys, based on [30]; (b) Lattice unit of MnGa with $L1_0$ (left) and $D0_{22}$ (right), arrows stand for magnetic moment directions of Mn atoms (Reprinted with permission from [17]).

### 2.2 Growth and magnetic properties of Mn$_x$Ga epitaxial films

In recent couple of decades, a great deal of effort has been made on growth of $L1_0$ and $D0_{22}$-Mn$_x$Ga films by magnetron sputtering or molecular-beam epitaxy (MBE). Since Krishnan and Tanaka et al. respectively observed square perpendicular hysteresis of Mn$_x$Ga ($1.2<x<1.5$) grown on GaAs (001) by magneto-optical, magnetic and transportation measurements in the early years of 1990s,[38,39] the growth and characterizations of $L1_0$-Mn$_x$Ga films have been intensively studied on many kinds of substrates, such as GaN, GaSb, Si, Al$_2$O$_3$ and MgO with different buffer layers (ScN and Cr).[36, 40-44] However, only a few of those films on GaAs and MgO demonstrated magnetic perpendicular anisotropy. $D0_{22}$-Mn$_x$Ga films have been prepared mainly on MgO (001) substrates.[22,35,36,45-47] Wu et al. reported magnetic and magneto-transport properties of $Mn_xGa$ (x=2.5 and 2.0) epitaxial films on Cr-MgO (001) and even found perpendicular anisotropy in 5 nm-thick films.[45,47] Kurt et al. prepared stoichiometric $D0_{22}$-$Mn_3Ga$ films with perpendicular easy axis on MgO, Pt-MgO and Cr-MgO, and observed the spin polarization of 58% (40%) in $Mn_3Ga$ ($Mn_2Ga$).[35,47] Recently, Nummy et al. and Zha et al. reported nanostructured $Mn_xGa$ films on Si/SiO$_2$ with coercivity over 20 kOe and discussed the possibility for permanent magnet applications, respectively.[48,49] Motivated by the great application potential and inspiring result of $Mn_xGa$ alloys, we prepared $Mn_xGa$ epitaxial films in a wide composition range (0.76≤ x ≤ 2.60) on GaAs (001) substrates by MBE, and carried out detailed studies on the tailoring of magnetism in these films with giant perpendicular anisotropy, ultrahigh coercivity and large energy product.[16,17]

Table 1. Lattice constants of representative metals, insulators and semiconductors and calculated misfit in respect to $D0_{22}$-$Mn_3Ga$ (a=3.910 Å), $L1_0$-MnGa (a=3.886 Å) and $L1_0$-MnAl (a=3.920 Å). * represents the lattice constants after 45° in-plane rotation ($a*=a/\sqrt{2}$).

| buffer | a (Å) | $f_{D022\text{-}Mn3Ga}$ (%) | $f_{L10\text{-}MnGa}$ (%) | $f_{L10\text{-}MnAl}$ (%) |
|---|---|---|---|---|
| Cu | 3.61 | 8.3 | 7.6 | 8.6 |
| Pd | 3.89 | 0.5 | -0.1 | 0.8 |
| Pt | 3.92 | -0.3 | -0.9 | 0 |
| Au | 4.08 | -4.2 | -4.8 | -3.9 |
| Ag | 4.09 | -4.4 | -5.0 | -4.2 |
| Al | 4.05 | -3.5 | -4.0 | -3.2 |
| Cr | 4.07* | -4.0 | -4.6 | -3.8 |
| Si | 3.94* | -0.8 | -1.4 | -0.5 |
| SrTiO$_3$ | 3.91 | 0 | 0.8 | 0.3 |
| GaAs | 4.00* | -2.1 | -2.7 | -1.9 |
| MgO | 4.21 | -7.2 | -7.8 | -7.0 |
| InAs | 4.28* | -8.6 | -9.1 | -8.4 |
| AlAs | 4.00* | -2.2 | -2.7 | -1.9 |

To realize the practical applications, especially for spintronics, it is essential to choose suitable substrates with small lattice misfit for epitaxial growth of $Mn_xGa$ films simultaneously with high magnetic performance, good structural quality and smooth surface. Table 1 summarizes the in-plane lattice constants of representative metals, insulators and semiconductors with (001) orientation and calculated misfit f of $D0_{22}$-$Mn_3Ga$ (a=3.910 Å) and $L1_0$-MnGa (a=3.886 Å). Simply in view of lattice misfit, Pt (001), Pd (001), Si (001), SrTiO$_3$ (001) and GaAs (001) should be the best choices for epitaxial growth of both $D0_{22}$ and $L1_0$ films. To our best knowledge, the report of both $L1_0$- and $D0_{22}$- $Mn_xGa$ epitaxial films grown on Pd is still lacking. Perpendicularly magnetized $L1_0$-$Mn_xGa$ was not available on Si (001) probably due to strong reaction between metal layer and active silicon though f between $L1_0$-MnGa and Si (001) is only 1.4%.[43] So far, only Pt, GaAs and SrTiO$_3$ buffers have been experimentally confirmed to be good choice for epitaxial growth of $Mn_xGa$ films with flat surface and high crystalline quality.[16, 17, 35, 47] As a recent review outlined,[24] the surfaces of $Mn_xGa$ films on Cr and MgO buffers appeared to be very rough with discontinuous islands,[47] whereas films on Pt and GaAs buffers exhibited very smooth surfaces, since the misfit of $D0_{22}$-$Mn_3Ga$ is -7.2%, -4.0%, -0.3% and 2.2% in respect to MgO, Cr, Pt and GaAs buffers, respectively. Therefore, it seems difficult to directly prepare smooth $Mn_xGa$ films on MgO or sharp $Mn_xGa$-MgO interface for high-TMR MTJ due to the large misfit.[50-53] Recently, Glas et al. also reported smooth $Mn_xGa$ (x=2.9) films with root mean square (RMS) even below 0.25 nm on SrTiO$_3$.[54] Zhu et al. reported $Mn_xGa$ samples (0.76≤x≤2.60) grown on GaAs with good homogeneity and abrupt interfaces[16,17] as revealed by the low-magnification TEM image in the inset of Fig. 2(a).

The structural and magnetic properties of $Mn_xGa$ films could be tailored effectively by several reliable methods, such as by controlling growth temperature ($T_s$),[16] tuning composition (x),[17, 36] changing annealing conditions [17] and selecting different substrates.[42,43] In the following paragraphs, we will show in detail how $T_s$ and x influence the structure and magnetism, taking $Mn_xGa$ films grown on GaAs (001) as examples.

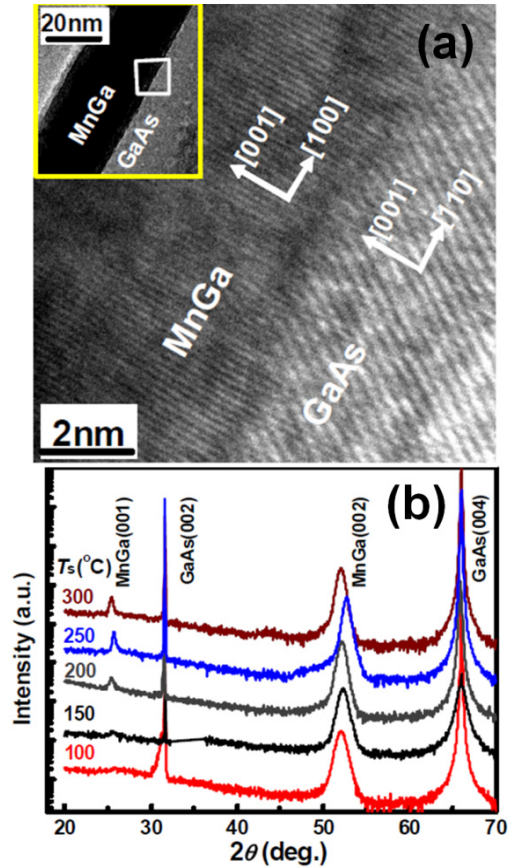

Fig. 2. (a) TEM image of MnGa films ($x=1.5$) grown on GaAs (001). The inset shows low-magnification TEM image. (b) XRD patterns of MnGa films ($x=1.5$) grown at 100, 150, 200, 250 and 300 $^{o}$C, respectively. Reprinted with permission from [16].

Figure 2(a) shows representatively a high-resolution cross-sectional transmission electron microscopy (TEM) image of $Mn_xGa$ films grown on GaAs,[16] indicating a well-textured $Mn_xGa$ layer with $c$-axis perpendicular to GaAs substrate. The epitaxial relationship of MnGa(001)[100]||GaAs(001)[110] can be derived from the high-resolution TEM image, coinciding with reflection high energy electron diffraction (RHEED) patterns. Figure 2(b) shows examples of synchrotron x-ray diffraction $\theta$-$2\theta$ patterns of $Mn_xGa$ ($x=1.5$) films grown at different temperature.[16] For all the films grown at temperatures from 100 to 350 $^{o}$C, only sharp (001) superlattice peaks and (002) fundamental peaks of $L1_0$-$Mn_xGa$ films can be observed in the range from 20$^{o}$ to 70$^{o}$ besides the peaks of GaAs substrates, indicating that these are all (001)-textured single-crystalline films with $c$-axis along the normal direction, in agreement with cross-sectional TEM images and *in situ* RHEED patterns.

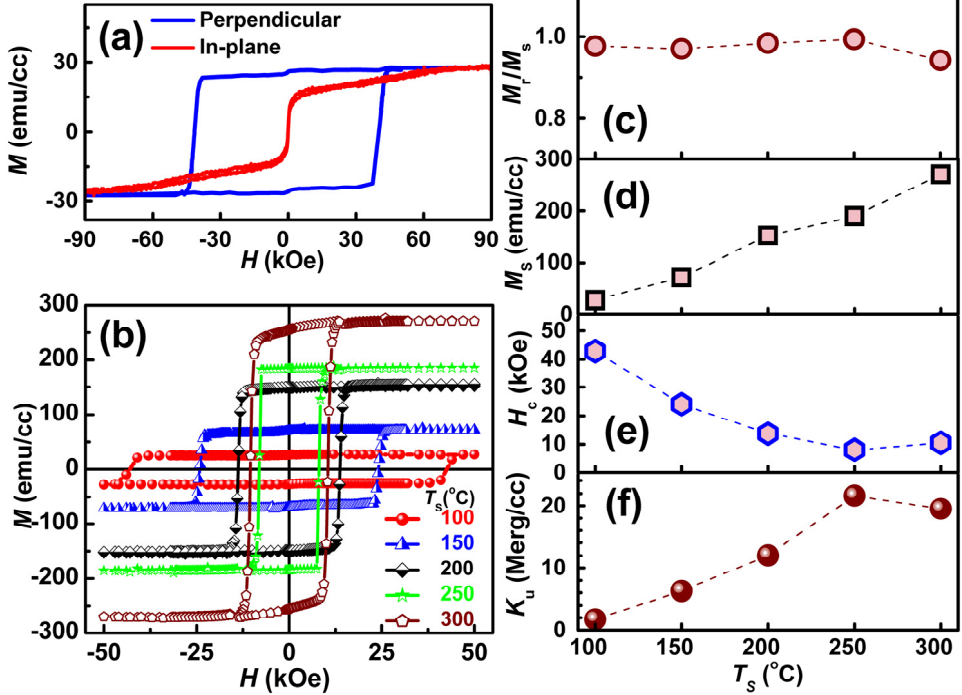

Fig.3. (a) Perpendicular and in-plane hysteresis loops of the 100 $^{o}$C-grown $L1_0$-$Mn_{1.5}$Ga films measured by PPMS. (b) Perpendicular hysteresis loops of the $L1_0$-$Mn_{1.5}$Ga films grown at different temperature ($T_s$) measured by SQUID. (c) $M_r/M_s$, (d) $M_s$, (e) $H_c$ and (f) $K_u$ plotted as a function of $T_s$. Reprinted with permission from [16].

Figure 3(a) shows both perpendicular and in-plane hysteresis loops of a typical $Mn_{1.5}$Ga film measured at 300 K.[16] A nearly square loop with ultrahigh coercivity ($H_c$) of 42.8 kOe and remnant magnetization ($M_r$) of 27.3 emu/cc is observed from the perpendicular $M$-$H$ curve. However, the in-plane $M$-$H$ curve exhibits almost anhysteretic loop, zero remnant magnetization and high saturation field exceeding 90 kOe. This observation offers a direct evidence for the giant perpendicular magnetic anisotropy. The same feature holds for all the films grown at temperatures between 100 and 300 $^{o}$C. Figure 3(b) shows the perpendicular $M$-$H$ curves of $Mn_{1.5}$Ga films grown at different temperature, revealing a significant influence of growth temperature ($T_s$) on the magnetic properties. Figures 3(c)-3(f) summarize the values of $M_r/M_s$, $M_s$, $H_c$ and $K_u$ as a function of $T_s$. As shown in Fig. 3(c), $M_r/M_s$ keeps a nearly invariant value large than 0.94 for all the films, very different from those of $L1_0$-$Mn_{1.5}$Ga films grown on GaN, GaSb, Si and $Al_2O_3$.[40-49] $M_s$ increases monotonically from 27.3 to 270.5 emu/cc as $T_s$ increases, which is much smaller than the calculated value of 2.51 $\mu_B$/Mn (~845 emu/cc) for stoichiometric $L1_0$-MnGa[16,21-23] due to both the overall strains and off-stoichiometry. As shown in Fig. 3(e), $H_c$ has a maximum of 42.8 kOe at 100 $^{o}$C, then drops to 8.1 kOe at 250 $^{o}$C, and finally climbs up to 10.8 kOe at 300 $^{o}$C. The ultrahigh $H_c$ in these $L1_0$-$Mn_{1.5}$Ga films was attributed to the combination of giant perpendicular magnetic anisotropy and imperfection including chemical disorder, lattice defects and overall tensile strains.[16] Importantly, the imperfection contribution to $H_c$ seems to be the same order of magnitude as the contribution from the magnetic anisotropy since $H_c$ can be tailored so obviously over the large scale from 8.1 to 42.8 kOe. With increasing $T_s$, $K_u$ goes up monotonically and reaches the maximum value at 250 $^{o}$C and then decreases with further increasing $T_s$. The maximum $K_u$ of 21.7 Merg/cc is roughly consistent with the theoretical value of 26 Merg/cc for $L1_0$-$Mn_{50}Ga_{50}$,[21] while much larger than reported $Mn_xGa$ films grown on other semiconductors[38-41] and MgO.[22] Importantly, such a giant $K_u$ of 21.7 Merg/cc in these $L1_0$-$Mn_{1.5}$Ga films supports nanoscale spintronic devices, such as spin-transfer-torque MRAMs bits, down to 5 nm in size and high-density bit-patterned recording with areal density up to 27 Tb/inch$^2$ and 60-year thermal stability.[16]

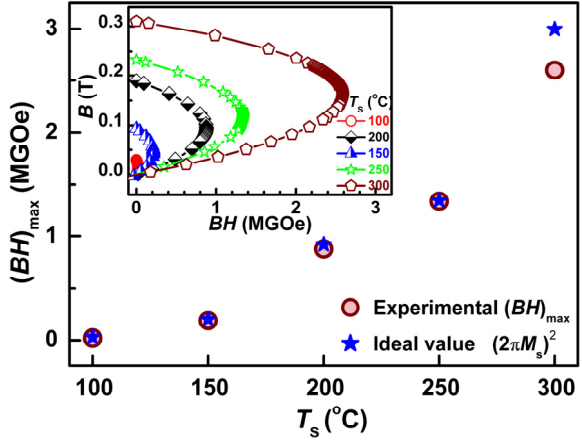

Fig. 4. Experimental magnetic energy products $(BH)_{max}$ of $L1_0$-$Mn_{1.5}$Ga films with comparison with ideal value $(2\pi M_s)^2$. The inset shows the $B$-$BH$ curves. Reprinted with permission from [16].

Magnetic materials with large magnetic energy product and noble-metal-free composition are also promising in permanent magnets applications.[20] The magnetic energy product $(BH)_{max}$ of $L1_0$-$Mn_xGa$ films was further evaluated and compared with the ideal value $(2\pi M_s)^2$ in Fig. 4. $(BH)_{max}$ climbs up monotonically from 0.02 to 2.60 MGOe as $T_s$ increases and almost the same as ideal value $(2\pi M_s)^2$ when $T_s<250$ °C because of the very high $M_r/M_s$. $(BH)_{max}$ at $T_s=300$ °C is slightly smaller than the ideal value probably due to the rounding at corner of $M$-$H$ loops and the small difference between $M_s$ and $M_r$ ($M_r/M_s = 0.94$). The largest magnetic energy product up to 2.60 MGOe are observed in the film grown at 300 °C, which is larger than that previously reported in bulk polycrystalline $D0_{22}$-$Mn_{2~3}$Ga alloys (<5.5 kJ/m³ or 0.68 MGOe)[35] and steels magnets (~1 MGOe)[20], and also comparable with hard ferrite magnets (~3 MGOe).[20] Therefore, the rare-earth-free compositions, high coercivity and large magnetic energy product together make this kind of $L1_0$-$Mn_xGa$ films promising to be developed for economical permanent magnets.

Both the structural and magnetic properties of $Mn_xGa$ epitaxial films were observed to be very sensitive to the composition. Recently, Mizukami *et al.* and Zhu *et al.* carried out symmetrical studies on the synthesis and strongly $x$-dependent magnetic properties of (001)-orientated $Mn_xGa$ epitaxial films with $x$=1.2-3 grown on Cr-buffered MgO (001)[36] and films with $x$=0.76-2.6 grown on GaAs (001),[17] respectively. Figure 5(a) shows examples of XRD $\theta$-$2\theta$ patterns of 250 °C-grown $Mn_xGa$ films with different $x$ grown on GaAs (001) by MBE.[17] For $0.76 \leqslant x \leqslant 1.75$, $L1_0$-MnGa films shows both sharp (001) superlattice peaks and (002) fundamental peaks in the range from 20° to 70° besides the peaks of GaAs substrates, indicating that these are all (001)-textured single-crystalline films. For film with $x$=0.74, three small peaks irrelative to $L1_0$ appear, which become dominating for film with $x = 0.55$. The new phase could be best fitted by cubic $MnGa_4$ with enlarged $c$ axis of 3.91 Å. For film with $x = 2.60$, only the fundamental peak could be found due to the high disorder. As shown in Fig. 5(b), the x-ray reflection (XRR) curves with the strong oscillations suggest sharp interfaces and good homogeneity of these samples.

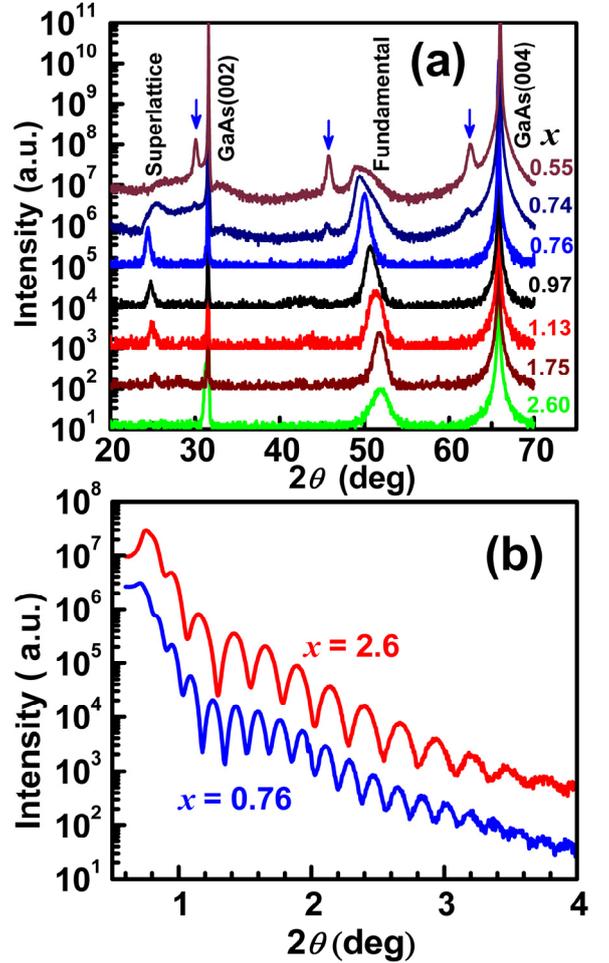

Fig. 5. (a) XRD patterns of MnGa films with different composition ($x$). (b) XRR curves of MnGa films with $x = 0.76$ and 2.6. Reprinted with permission from [17].

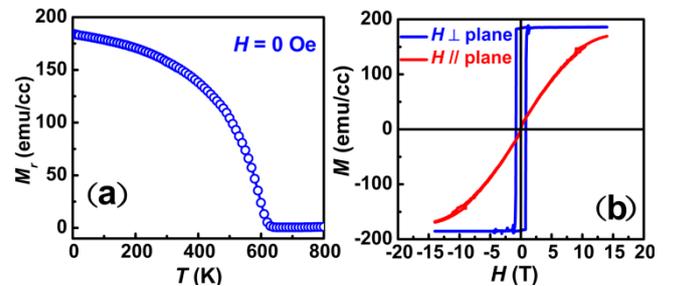

Fig. 6. (a) Temperature dependence of remanent magnetization ($M_r$); (b) Perpendicular and in-plane hysteresis loops of $Mn_{1.4}$Ga films grown at 250 °C. Reprinted with permission from [17].

L J Zhu et al. ArXiv:1309.0298 (2013)5

Figure 6(a) shows a typical temperature dependence of remanent magnetization of $Mn_xGa$ films with Curie temperature of 630 K.[17] Figure 6(b) show an example of hysteresis loops of $L1_0$-$Mn_{1.4}Ga$ films grown at 250 °C.[17] The sample exhibits a fairly square perpendicular loop with high perpendicular coercivity of 11 kOe and anhysteretic in-plane loops with giant saturation fields exceeding 14 T. These characteristics imply strong anisotropy existing in this kind of films.

Figures 7(a)-(c) display the $x$-dependent magnetic properties including $M_r/M_s$, $M_s$ and $H_c$ determined from the 300 K perpendicular $M$-$H$ curves of these 250 °C-grown films. The $Mn_xGa$ films with $x$=0.97-1.75 exhibit high $M_r/M_s$ exceeding 0.90; but those with $x$ deviated from this range show dramatically decreased $M_r/M_s$. As shown in Fig. 7(b), $M_s$ decreases dramatically from 450 to 52 emu/cc with $x$ increases from 0.76 to 2.60, which should be attributed partly to the increase of antiferromagnetic coupling between Mn atoms at different sites,[21] partly to the increase of strains induced by short $c$ axis.[16,23] The maximum $M_s$ of 445 emu/cc in $Mn_{0.76}Ga$, is still below the calculated value of 845 emu/cc for stoichiometric $L1_0$-MnGa[16,21] and experimental value of 600 emu/cc in 400-450 °C-annealed $Mn_{1.17}Ga$ grown on MgO[36] probably due to the low growth temperature. $H_c$ climbs up from 4.38 to 20.1 kOe as $x$ increases from 0.76 to 1.75, and then drops to 1.1 kOe at $x$=2.60. This change tendency is consistent well with that of $K_u$ estimated in Fig. 7(d) following the method used in ref. 16. $K_u$ exhibits large values from 8.6 to 21.0 Merg/cc in $L1_0$ range, while quickly degrades to 0.02 Merg/cc in $D0_{22}$-$Mn_{2.6}Ga$ grown on GaAs because of increased disorder. Noticeably, $D0_{22}$-$Mn_xGa$ with high ordering and high quality may also demonstrate high $K_u$ values, such as those grown on MgO substrates at high temperatures.[36,45] As ever discussed,[16,25] rare-earth-free and noble-metal-free $Mn_xGa$ alloys are considerable for economical permanent magnet application. Since intrinsic coercivity $H_c$ determined from $M$-$H$ curves, normal coercivity $H_{cB}$ determined from $B$-$H$ curves and magnetic energy products $(BH)_{max}$ are three key features of quality of permanent magnets, we further calculated $H_{cB}$ and $(BH)_{max}$ in Figs. 7(e) and (f). With increasing $x$ in the phase-pure range, both $H_{cB}$ and $(BH)_{max}$ decrease nearly monotonically, corresponding to the decrease of $M_s$. The film with $x$ = 1.07 shows the highest $H_{cB}$ and $(BH)_{max}$ of 3.6 kOe and 3.4 MGOe, respectively. The maximum of 3.4 MGOe is larger than that recent reported in $Mn_{1.5}Ga$ film (2.6 MGOe)[16] and ferrite magnets (3 MGOe),[20] which makes $L1_0$-$Mn_xGa$ alloys with low Mn compositions still promising to be developed for cost-effective and high-performance permanent magnets applications. Taking into consideration of all the magnetic properties shown in Figs. 7(a)-7(f), low-Mn-composition $L1_0$-$Mn_xGa$ films are more advantageous than those $D0_{22}$ films to be developed for applications in magnetic recording with areal density over 10 Tb/inch$^2$, high-performance nanoscale spintronic devices like MRAM bits with size down to several nm and economical permanent magnet applications, since they simultaneously exhibit high $M_r/M_s$, $M_s$, $K_u$, $H_c$, $H_{cB}$ and $(BH)_{max}$.

For further understanding of the composition-dependent magnetism, the structural and magnetic data for typical $Mn_xGa$ films grown on MgO (001) substrates are summarized in Table 2. The magnetization of these films grown on MgO substrates degrades monotonically with increasing $x$, in consistence with on the case of $Mn_xGa$ films grown on GaAs, which should be attributed partly to the increase of antiferromagnetic coupling between Mn atoms at different sites,[21] partly to the increase of strains induced by short $c$ axis.[16,23] These films also demonstrated giant perpendicular anisotropy $K_u$ of $10^6$-$10^7$ erg/cc and huge coercivity $H_c$ varying from 1 to 18 kOe. Noticeably, the values of $H_c$ always tend to be larger in films with higher Mn-composition than films with lower Mn-composition, while values of $K_u$ seem to have complicated dependences on compositions, probably due to the influence of different growth conditions or buffers.

It is also worth mentioning that the magnetism could be effectively tailored by post-growth annealing [17] and choosing different substrates. By using different substrates like semiconductors (GaAs, GaSb, GaN, Si, ScN, etc), insulators ($SiO_2$, $Al_2O_3$, MgO, etc) and metals (Cr, Pt, etc), $Mn_xGa$ films would crystallized into different structures, such as different crystalline orientation,[42,43] different morphology [35,47-49] and different strains,[16] and further have different magnetic anisotropy and other magnetic properties.

Therefore, $Mn_xGa$ films are observed to exhibit giant perpendicular magnetic anisotropy, flexible magnetization, huge coercivity, large energy products and high Curie temperature. Meanwhile, the magnetism of $Mn_xGa$ films could be effectively tailored via several reliable ways, especially by controlling growth temperature, tuning compositions, post-growth annealing and selecting substrates. These data provide a comprehensive diagram for tailoring and improving the structure and magnetism of $Mn_xGa$ films, which would be helpful for understanding this kind of material and approaching successful applications in a variety of spintronic devices, magnetic recording and permanent magnets

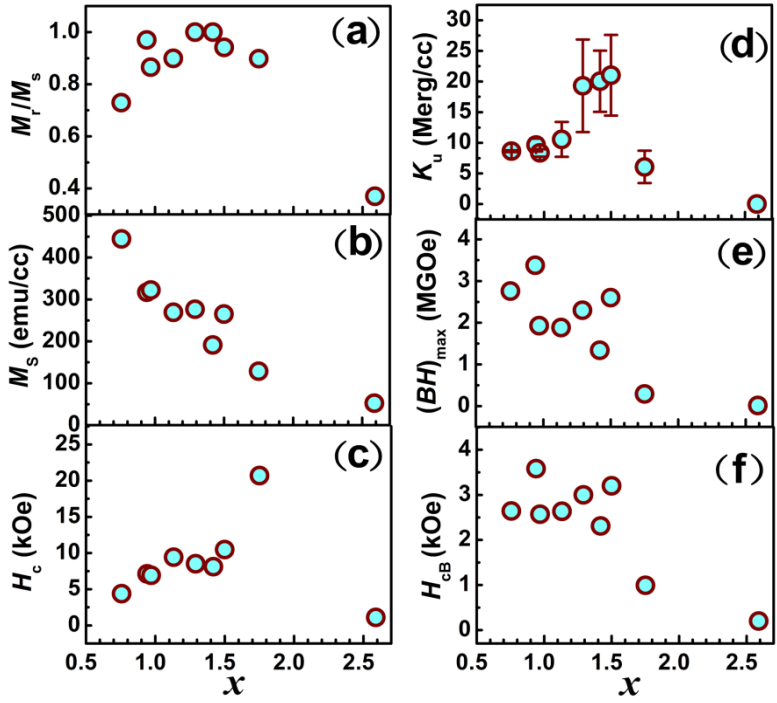

Fig. 7. Composition $x$ dependence of (a) $M_r/M_s$, (b) $M_s$, (c) $H_c$ and (d) $K_u$ of $Mn_xGa$ films grown on GaAs (001) at 250 $^oC$. Reprinted with permission from [17].

Table 2 Structural parameters and magnetic data of typical $Mn_xGa$ films grown on MgO (001). $T$ represents growth or annealing temperature.

|  | Substrate | $T$ ($^oC$) | a (Å) | c (Å) | $M_s$ (300 K) (emu/cc) | $K_u$(300 K) (Merg/cc) | $H_c$ (300 K) (kOe) | Reference |
|---|---|---|---|---|---|---|---|---|
| $Mn_3Ga$ | Pt-MgO | 250 | - | 7.12 | 110 | 8.9 | 18 | [35, 47] |
| $Mn_3Ga$ | MgO | 350 | - | 7.07 | 140 | 19.1 | 18 | [35, 47] |
| $Mn_3Ga$ | Cr-MgO | 350 | - | 6.96 | 140 | 4 | 17 | [35, 47] |
| $Mn_3Ga$ | Cr-MgO | 400 | - | 7.10 | 200 | - | 15 | [36] |
| $Mn_{2.5}Ga$ | Cr-MgO | 400 | 3.91 | 7.11 | 250 | 12 | 6 | [46] |
| $Mn_{2.1}Ga$ | MgO | 450-500 | 3.92 | 7.11 | 305 | 15 | - | [22] |
| $Mn_2Ga$ | MgO | 360 | - | 7.16 | 470 | 23.5 | 3.6 | [35] |
| $Mn_{1.2}Ga$ | Cr-MgO | 400-500 | 3.90 | 3.60 | 600 | 15 | 1 | [36] |

### 2.3 Other spin-dependent properties of $Mn_xGa$ epitaxial films
#### 2.3.1 Spin polarization

For magnetic sensors, memories and oscillators, it is advantageous to use materials with high spin polarization, which allows for high magnetoresistance ratio and high signal to noise ratio. As calculated by Sakuma and Winterlik et al. stoichiometric $L1_0$-MnGa and $D0_{22}$-$Mn_3Ga$ should have high spin polarization of 71% and 88% at the Fermi level, respectively.[21,25] The spin polarization was also predicted to decrease with Mn deficiency, atomic disorder and strain.[25] Kurt et al. performed point contact Andreev reflection spectroscopy measurements and determined spin polarization of 58% and 40% at 2.2 K in $D0_{22}$-$Mn_3Ga$ and $Mn_2Ga$ films, respectively.[47] Higher spin polarization was expected in perfectly ordered $D0_{22}$-$Mn_xGa$ films although detailed investigations are still missing. To our best knowledge, there have also been no available reports about experimental measurements of spin polarization of $L1_0$-$Mn_xGa$ alloys. Further investigations on spin polarization are needed.

#### 2.3.2 Magnetic damping

Spin-transfer-torque (STT) is a phenomenon of magnetization switching induced by spin current. For STT-driven spintronic devices, a low critical current for current-induced magnetization switching is preferred for low energy consumption and reliable switching.[2,55] The critical switching current can be written as:

$$I_{C0} = \alpha K_u V \frac{\gamma e}{\mu_B g(P,\theta)} \quad (1)$$

where $\alpha$ is magnetic damping constant; $K_u$, perpendicular anisotropy; $V$, volume of magnets; $\gamma$, gyromagnetic ratio; $\mu_B$, Bohr magnetron; $e$, the elementary charge; $g(P,\theta)$, function of spin polarization ($P$) of tunnel current and the angle ($\theta$) between the magnetization of the free and the reference layers.

As indicated in Eq. (1), perpendicular magnetized materials with combination of high $P$, low $\alpha$ and low $K_u$ are preferred for low-switching-current STT device applications. Generally, a high $K_u$ and a low critical switching current $I_{c0}$ are in contrast, but high $K_u$ is absolutely essential for long time stability. However, a combination of high $P$ and $K_u$ with low enough $\alpha$ could still provide a possibility for a low $I_{c0}$ and long time thermal stability. Except for the high $P$ and $K_u$, the noble-metal-free $Mn_xGa$ alloys also have much lower magnetic damping constants than those of all the other perpendicular-anisotropy materials to our best knowledge, due to the low density of states at the Fermi level and the weak spin-orbit interaction in light 3$d$ elements Mn and Ga.[22] Time-resolved MOKE experiment revealed the damping constants of 0.008 (0.015) for $Mn_{1.54}Ga$ ($Mn_{2.12}Ga$) films, which are one or two order smaller than that of known materials with large perpendicular anisotropy.[2] First-principles calculations indicate even lower damping constant of 0.0003 (0.001) for $L1_0$-MnGa ($D0_{22}$-$Mn_3Ga$) defect-free alloys.[22] Therefore, with respect to both high thermal stability and low critical switching current, $Mn_xGa$ alloys with either $L1_0$ or $D0_{22}$ ordering are advantageous for low-current-switching STT-memory and oscillator applications.

### 2.3.3 Magneto-optical Kerr effect

Perpendicularly magnetized materials with large Kerr rotation angle and optical reflectivity are also promising in high-density perpendicular magneto-optical recording. Krishnan first studied the magneto-optical properties of $L1_0$-$Mn_{1.5}Ga$ films grown on GaAs (001) in 1992.[38] The film exhibited a large Kerr rotation angle of ~0.1° at 820 nm. Furthermore, reflectivity around 65%-70% were found over a broad wavelength range from ultraviolet to far infrared. From a technological point of view, the epitaxial growth on semiconductor and the high-performance magneto-optical properties would make the exciting possibility of the integration of magneto-optic and semiconductor devices a reality.

### 2.3.4 Anomalous Hall effect

Anomalous Hall effect (AHE), related to spin-orbit interaction in magnetic materials, has potential for sensor, memory and magnetic logic applications.[45] Large AHE has been observed in various PMA films including $L1_0$-FePt, FePd, etc.[56,57] Accordingly, study of AHE in $Mn_xGa$ films is of great importance not only for practical applications but also helpful for comprehending the widely disputed mechanism of AHE. Tanaka et al. measured the AHE hysteresis of $L1_0$-$Mn_{1.5}Ga$ films on GaAs and found the AHE resistivity sample-dependent, changing from 0.5-4 μΩ.cm.[39] Bedoya-Pinto revealed noticeably composition-dependent scaling behavior in $L1_0$-$Mn_xGa$ ($x$=0.96, 1.38 and 2.03) films on GaN.[41] Wu et al. observed AHE resistivity up to 11.5 μΩ.cm in $D0_{22}$-$Mn_2Ga$,[45] which is much larger than that of other perpendicular-anisotropy materials, such as $L1_0$-FePt (1-6 μΩ cm).[56,57] Glas et al. studied $D0_{22}$-$Mn_xGa$ ($x$=2.3, 2.6 and 2.9) films grown on MgO and $SrTiO_3$, and discussed the determined skew scattering and side jump coefficients with regard to the film composition and compared to the crystallographic and magnetic properties.[54] Zhu et al. reported AHE and the scaling behavior of $L1_0$-$Mn_{1.5}Ga$ films grown on GaAs (001) and found significant influence of disorder on the scaling behaviors.[58] However, for further understanding of the underlying physics, more detailed studies will be needed.

### 2.4. Thermal and Chemical stability

Structural stability of materials is of particular importance for practical applications. $L1_0$-ordered $Mn_xGa$ have been widely believed to keep thermal-dynamically stable up to at least 800 K,[31,32] by contrast, $Mn_xGa$ with $D0_{22}$-ordering are observed to be metastable and will transfer to antiferromagnetic hexagonal $D0_{19}$ phase at temperature over 770 K.[25,33,34] Zhu et al. investigated the thermal stability of MnGa/GaAs interface by performing systematical post-growth annealing experiments taking $L1_0$-$Mn_{0.76}Ga$ film as an example, which exhibits the largest $M_s$ among all the as-grown films with different $x$ (Fig. 7(b)). Figure 8(a) shows XRD patterns of as-grown and 10 min-annealed $Mn_{0.76}Ga$ films, from which we found the $Mn_{0.76}Ga$ film itself is thermal stable up to 450 °C, and the interface could keep stable up to at least 350 °C. This result roughly agrees with previous report that $Mn_xGa$ with $x$<1.5 should keep stable on GaAs below 400 °C.[59] Interestingly, very weak $Mn_2As$ (003) peak as a result of interfacial reaction is observable by high-sensitivity synchrotron XRD measurement, which may be undetectable by common XRD with Cu K$\alpha$ radiation.[59] Fortunately, the stable temperature of at least 350 °C could still satisfy the demand of integration with metal-oxide-semiconductor transistors.

As to the chemical stability in the ambient conditions, there have been disputations on the chemical stability of $Mn_xGa$ alloys. Roy et al. found an oxide layer of less than 5 nm formed at the $Mn_xGa$ surface,[60] but Wang et al. found no signs of oxidation of $Mn_xGa$ films in air atmosphere.[61] In fact, a capping layer is always used to prevent possible oxidation of $Mn_xGa$ film.[16,36,38] In order to make clear the chemical stability in the ambient conditions, we show in Fig. 8(b) x-ray photoelectron spectroscopy (XPS) of Mn and Ga on the surfaces of $Mn_xGa$ films with and without capping layer which were exposed in the air at room temperature for a week.[24] $Mn_xGa$ films with Al capping layer showed asymmetric elemental peaks for both Mn 2$p$ and Ga 3$d$ spectrums. The asymmetry of these peaks should be attributed to the significant shielding effect to the core levels by the high-density at the Fermi level due to the alloy nature of the film, which further reveals the good protection of a 2 nm Al layer from oxidation. By contrast, $Mn_xGa$ films without Al capping layer showed evident peak shifts and reduced asymmetry, indicating significant oxidation occurred. Therefore, it is necessary to protect $Mn_xGa$ films from oxidation by using suitable capping layers.

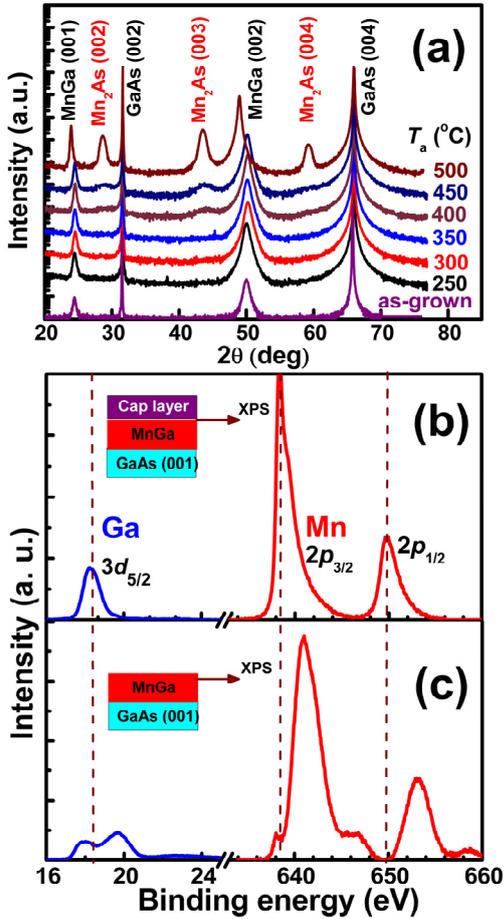

Fig. 8 (a) XRD patterns of $Mn_{0.76}Ga$ films annealed at 250, 300, 350, 400, 450 °C, respectively (Reprinted with permission from [17]); XPS patterns of $Mn_{0.97}Ga$ films (b) with and (c) without Al cap layer after exposed in air atmosphere for a week (Reprinted with permission from [24]).

### 2.5. $Mn_xGa$-based spintronic devices

Magnetic tunneling junctions (MTJ) and giant magnetoresistance (GMR) devices based on perpendicularly magnetized materials are expected to ensure thermal stability of nanoscale spintronic devices and to be developed as high-performance MR sensors, STT-MRAM and STT-oscillators. As discussed above, (001)-orientated $Mn_xGa$ alloys with $L1_0$ or $D0_{22}$ ordering simultaneously show giant $K_u$, high $P$, high $T_c$, huge $H_c$, moderate $M_s$ and low $α$, which make this kind of material an ideal combination for high-field sensors, Gbit STT-MRAM and high-power STT-oscillators. Recently, a few groups have carried out theoretical and experimental work from this point of view. Using first-principle calculations based on density functional theory and non-equilibrium Green's function, Bai $et$ $al.$ predicted large tunnel magnetoresistance (TMR) effect of 1000% in $Mn_2Ga/MgO/Mn_2Ga$ magnetic tunnel junctions, whereas no tunneling magnetoresistance effect in the $Mn_3Ga/MgO/Mn_3Ga$ MTJ stack due to the symmetry selective spin filtering effect of MgO.[62] Kubota $et$ $al.$ observed TMR ratios no more than 23% at 10 K and annealing endurance up to 375 °C in a series of $Mn_xGa/MgO/CoFe$ MTJ with different compositions of $Mn_xGa$ electrode.[50] The small TMR ratios were attributed to the large misfit between MgO and $Mn_xGa$, which may cause lattice dislocations and an unfavorable growth of the MgO barrier. Kubota $et$ $al.$ further made attempts to improve TMR ratios by using Fe or Mg insertion layers between $Mn_xGa$ and MgO interface,[51,52] however, the optimized TMR ratios still below 25% at 10 K. So far, the highest TMR ratio in $Mn_xGa$-based MTJs is 40% at room temperature observed in $Mn_{1.63}Ga/MgO/CoFeB$ MTJ with Co (1.5 nm)/Mg (0.4 nm) insertion between $Mn_{1.63}Ga$ and MgO.[53] Noticeably, Zha $et$ $al.$ also fabricated (112)-textured $Mn_{2.3-2.4}Ga/Cu/CoFe$ current-in-plane pseudo spin valve with GMR up to 3.88%.[63] Roy $et$ $al.$ studied the oscillatory dependence of the interlayer exchange coupling strength on spacer layer thickness in $Mn_xGa/GaAs$ ($Mn_2As$, $Mn_2Sb$)/$Mn_xGa$ trilayers.[64] In addition, Adelmann $et$ $al.$ demonstrated electrical spin-injection from $Mn_{1.38}Ga$ into (Al,Ga)As $p$-$i$-$n$ light-emitting diode (LED) at remanence.[12]

It is still an open issue to explore $Mn_xGa$-based MTJ, spin valve devices and semiconductor spintronic devices with high performance for practical applications, and take advantage of the amazing properties of this candidate material.

### 3. $L1_0$-$Mn_xAl$ epitaxial films on semiconductors
### 3.1 Lattice structure and synthesis of $Mn_xAl$ bulks

Despite the complex phase diagram (Fig. 9),[65] alloys of Mn and Al only have one ferromagnetic phase with a lattice unit similar to that of $L1_0$-MnGa (i.e. τ-phase) shown in Fig. 1(b). The lattice constants of $L1_0$-MnAl bulk are $a$=3.92 Å and $c$=3.57 Å.[66] In 1960, Koch $et$ $al.$ synthesized $L1_0$-$Mn_{1.25}Al$ bulk by rapid cooling of nonmagnetic ε-MnAl and observed $T_c$ of 653 K, $M_s$ of 490 emu/cc, $K_u$ of ~10 Merg/cc and $(BH)_{max}$ of 3.5 MGOe.[66] In 1977, Ohtani *et al.* synthesized MnAl bulk with giant $(BH)_{max}$ over 7 MGOe by doping C element into the alloys.[67] As shown in Fig. 9, a composition range with $x$ from 1 to 1.5 and annealing temperature between 500 to 900 °C are needed for synthesis of $L1_0$-MnAl bulks. $L1_0$-MnAl bulks are metastable phase and easily decompose into $\beta$-Mn and nonmagnetic $\gamma$-MnAl.

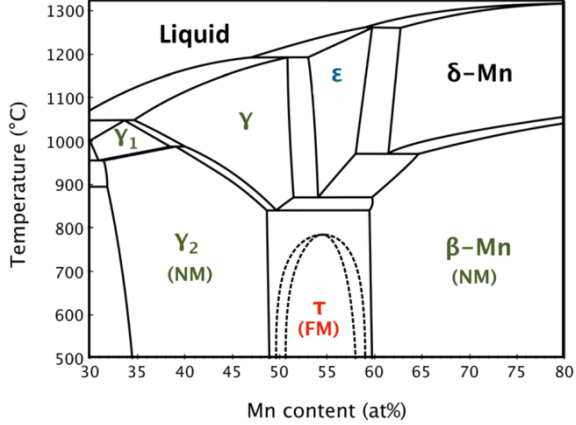

Fig. 9 Binary phase diagram of MnAl alloys. Reprinted with permission from [65].

### 3.2 Growth and magnetic properties of $Mn_xAl$ epitaxial films

Since 1990s, the growth and magnetic properties of $L1_0$-MnAl thin films have been widely studied on AlAs-buffered GaAs (001),[68,69,71-74] GaAs (001),[26,70] Cr-buffered MgO (001)[75,76] and glass[77] by MBE, sputtering, e-beam evaporator and pulse laser deposition. Despite of these progresses, only a few $L1_0$-MnAl epitaxial films grown on GaAs (001) and MgO (001) substrates exhibited perpendicular anisotropy.[68-76] We should point out that many metal (e.g. Pt, Pd, Al and Cr, etc), insulator (e.g. $SrTiO_3$) and semiconductor (e.g. Si, GaAs, InAs and AlAs) buffers could allow for epitaxial growth of $L1_0$-MnAl thin films in view of lattice mismatch (Table 1). However, there are no available reports on the epitaxial growth of $L1_0$-MnAl thin films on buffers or substrates other than GaAs (with or without AlAs buffer) and Cr-buffered MgO (001).

The magnetic properties of $L1_0$-MnAl thin films are observed to be very sensitive to the composition and growth temperature. As recently investigated systematically,[70] although $L1_0$-ordered $Mn_xAl$ films could been epitaxially grown on GaAs (001) in a wide composition range with $x$ from 0.5 to 1.2, films with $x=1.1$ exhibit the strongest XRD satellite peaks, indicating the optimized composition and the nearly perfect $L1_0$-ordered superlattice structures. $Mn_{1.1}Al$ (or $MnAl_{0.9}$) also shows the highest $M_r/M_s$ and $M_s$.[70] For further optimizing and tailoring of the structure and magnetic properties the systematical experiments were carried out on the growth temperature effects.[26] As shown in Fig. 10, the XRD patterns for $Mn_{1.1}Al$ films grown at different $T$s from 100 to 400 °C, revealing that all the $Mn_{1.1}Al$ films are single-crystalline with $c$-axis along the normal direction. Moreover, $Mn_{1.1}Al$ film at 350 °C exhibits the satellite peaks disclosing the optimized growth temperature range for the best crystalline quality. However, $Mn_{1.1}Al$ films grown beyond 250 °C show a peak of nonmagnetic $\varepsilon$-phase. The sharp interface between $Mn_{1.1}Al$ and GaAs were confirmed by both the high-resolution TEM image and XRR curves.

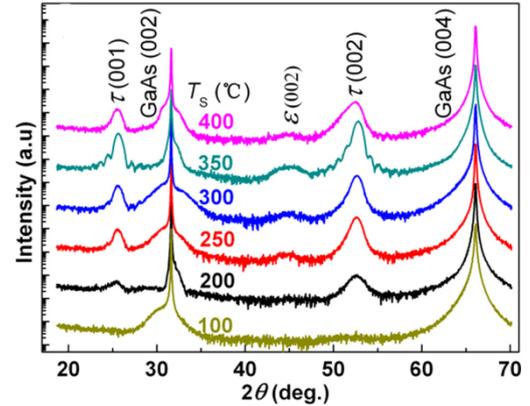

Fig. 10. XRD patterns of $L1_0$-$Mn_{1.1}Al$ films grown at 100, 200, 250, 300, 350 and 400 °C, respectively. Reprinted with permission from [26].

Figure 11(a) shows typical perpendicular and in-plane hysteresis loops of $L1_0$-$Mn_{1.1}Al$ sample grown at 350 °C. The almost square perpendicular hysteresis loop and anhysteretic in-plane hysteresis loop clearly demonstrate that the sample is more easily magnetized in out-of-plane direction. The perpendicular hysteresis loops of MnAl films grown at different $T_s$ are shown in Fig. 11(b), and it can be found that perpendicular $M_r/M_s$, $M_s$, $H_c$ and $K_u$ strongly depend on $T_s$, the details of which are summarized in Figs. 11(c) and 11(d). $M_s$ increases from 47.8 to 361.4 emu/cc as $T_s$ increases from 100 to 350 °C, and then decreases to 316.4 emu/cc at 400 °C. These changes are mainly linked to crystalline quality of the samples. $H_c$ has a minimum of 330 Oe at 100 °C, then increases to 10.7 kOe at 250 °C, and finally drops to 8.1 kOe at 400 °C. $K_u$ show a maximum of 15 Merg/cc at 350 °C. Moreover, MnAl films are another candidate for rare-earth-free permanent magnets. As shown in Fig. 11(a), $(BH)_{max}$ first increases from 0.001 to 4.44 MGOe as $T_s$ increases from 100 to 350 °C, then decreases to 2.46 MGOe at 400 °C. As compared in Fig. 11(b), the experimental value of $(BH)_{max}$ is lower than the ideal value $(2\pi M_s)^2$ probably due to the differences between $M_s$ and $M_r$ and the rounding of $M$-$H$ loops. The largest $(BH)_{max}$ up to 4.44 MGOe is larger than that previously reported in MnAl alloys (3.5 MGOe),[66] $Mn_{1.5}Ga$ films (2.6 MGOe),[16] and hard ferrite magnets (3MGOe).[20]

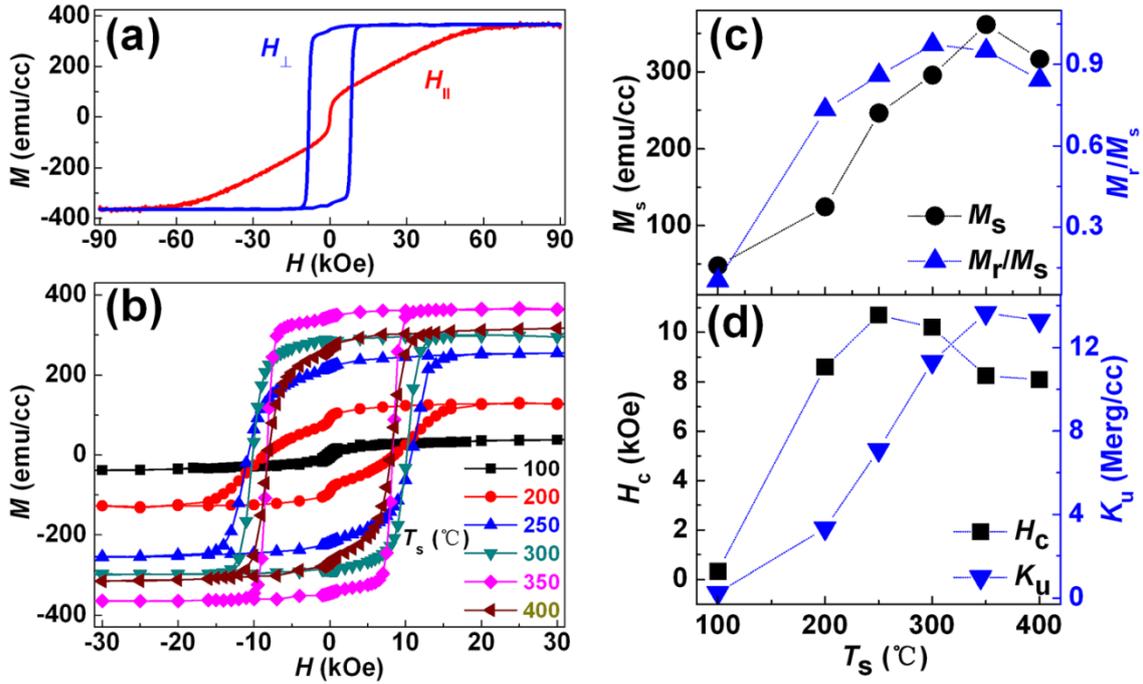

Fig. 11. (a) Perpendicular and in-plane hysteresis loops of the 350 °C-grown $L1_0$-Mn$_{1.1}$Al films. (b) Perpendicular hysteresis loops of the $L1_0$-Mn$_{1.1}$Al films grown at different temperature ($T_s$). (c) $M_r/M_s$, $M_s$, (d) $H_c$ and $K_u$ plotted as a function of $T_s$. Reprinted with permission from [26].

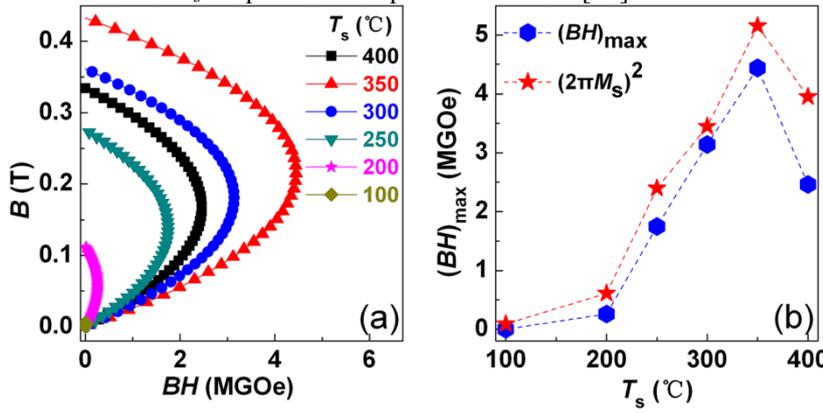

Fig. 12 (a) $T_s$ dependence of the magnetic energy products. (b) Comparison of the experimental $(BH)_{max}$ with ideal value $(2\pi M_s)^2$. Reprinted with permission from [26].

For better understanding, the typical data of structure and magnetic properties of MnAl films grown on different buffers are compared in Table 3. Mn$_x$Al films grown on AlAs and GaAs buffers exhibited much lower magnetization than bulk and films grown on Cr-MgO, which is probably related to the shorter $c$-axis and corresponding strains. Noticeably, despite the different buffers and growth (annealing) temperature, all these MnAl films display a high $K_u$ exceeding 10 Merg/cc.

Table 3 Structure parameters and magnetic data for representative Mn$_x$Al films grown on different substrates. $T_s$ ($T_a$) represents growth (annealing) temperature.

| | Substrate | $T_s$ ($T_a$) (°C) | $a$ (Å) | $c$ (Å) | $M_s$ (300 K) (emu/cc) | $K_u$ (300 K) (Merg/cc) | $H_c$ (300 K) (kOe) | Reference |
|---|---|---|---|---|---|---|---|---|
| Mn$_{1.50}$Al | AlAs-GaAs | 200~270 (350) | - | 3.41 | 285 | - | ~4.0 | [73] |
| Mn$_{1.11}$Ga | GaAs | 350 | - | 3.47 | 361 | 13.7 | ~8.0 | [26] |
| Mn$_{1.32}$Al | Cr-MgO | 250 (250) | - | ~3.57 | 530 | 10 | - | [75] |
| Mn$_{0.92}$Al | Cr-MgO | 200 (450) | 3.92 | 3.57 | 600 | 10 | - | [76] |

### 3.3 Other spin-dependent properties
#### 3.3.1 Magneto-optical Kerr effect

Cheeks *et al.* investigated the polar magneto-optical Kerr effect in 5 and 10 nm thick $L1_0$-MnAl epitaxial films grown on AlAs-buffered GaAs (001) and found a large Kerr rotation of 0.11° in a wide wave length range from 220 to 820 nm.[71] The 0.11° value is modest compared to the 0.3-0.4° values reported for materials such as TbFeCo, currently used in the industry. Although thicker films are expected to give higher Kerr rotation, there is so far no further progress report available.

#### 3.3.2 Anomalous Hall effect

The anomalous Hall effect of $L1_0$-MnAl films were intensively studied in 1990s. Sands *et al.*[68] and Leadbeater *et al.*[69] prepared single-crystalline $L1_0$-MnAl films on AlAs-GaAs (001) with a perpendicular *c*-axis and observed a large Hall hysteresis. Angadi and Thanigaimani reported the Hall effect of MnAl films evaporated on glass substrates with different compositions and thickness (25-150 nm), and found that the ordinary Hall coefficient $R_0$ and anomalous Hall coefficient $R_s$ are of the same order over the entire thickness range.[77] Both $R_0$ and $R_s$ are positive for lower thickness, and negative for higher thickness. Boeck *et al.* observed rectangular Hall hysteresis loops in $Mn_{1.5}Al$ grown on AlAs-GaAs (001) and AHE resistivity in the order of 4-8 μΩ.cm,[72] depending on width of Hall bars. Despite of these progresses, detailed knowledge about the microscopic mechanism of AHE in this kind of material remains a shortage and needs further exploration in the future.

### 4. Conclusions and outlook

Numerous studies have been carried out on the synthesis and characterizations of $Mn_xGa$ and $Mn_xAl$ epitaxial films with perpendicular anisotropy. $Mn_xGa$ alloys with both $L1_0$ and $D0_{22}$ ordering show giant perpendicular anisotropy, high spin polarization, high Curie temperature, huge coercivity, moderate magnetization, large energy product, low damping constant, giant Kerr rotation and large anomalous Hall resistivity and high thermal dynamical stability. $L1_0$–ordered $Mn_xAl$ films grown on GaAs and MgO also exhibit giant perpendicular anisotropy, high Curie temperature, huge coercivity, moderate magnetization, large energy product, giant Kerr rotation and large anomalous Hall resistivity (2-8 μΩ.cm). Importantly, the magnetism of both $Mn_xGa$ and $Mn_xAl$ films could be tailored effectively by several methods including controlling growth temperature, tuning composition, varying annealing conditions and selecting different substrates. These pronounced magnetic properties and effective controllability make these perpendicularly magnetized Mn-based binary alloy films much promising for high-performance novel spintronic devices, ultrahigh-density perpendicular magnetic recording and economical permanent magnets, although there have rarely been progress on high-performance spintronic devices, magnetic recording or practical permanent magnets applications. Moreover, the $Mn_xGa$ and $Mn_xAl$ ferromagnetic films with both high $K_u$ and good compatibility with semiconductor have great applications in not only semiconductor spintronics like spin-LED, spin-FET, and lateral spin valves with perpendicular spin injectors and detectors，but also in high-density integration of metallic spintronics functional devices like nonvolatile magneto-resistive random access memory on semiconductor photonic and electronic circuits.

We expect more detailed exploration on fundamental issues as to detailed knowledge on spin polarization, magnetic damping and AHE of $Mn_xGa$ and $Mn_xAl$ alloys, and possible practical applications like ultrahigh-density magnetic recording, economical permanent magnets, nonvolatile memory, high-field magnetoresistive sensors and high-performance oscillators and semiconductor spintronic devices with perpendicular injectors and detectors taking advantage of the amazing properties of this candidate material. It is also admirable to explore new kinds of ferromagnetic films with both high $K_u$ and good compatibility with semiconductors. We are also looking forward to new progress on high-density integration of metallic spintronics functional devices with semiconductor circuits taking advantage of such films.